\documentclass[10pt,letterpaper,twocolumn]{article}
\usepackage{ol2}
\usepackage{hyperref}
\usepackage{amsmath,amssymb,graphicx}
\usepackage{comment}
\usepackage{xcolor}
\usepackage{stfloats}
\usepackage{cleveref}

\crefname{equation}{}{}

\begin{document}

\twocolumn[
\title{Add-drop filter based on dual photonic crystal\\nanobeam cavities in push-pull mode}

\author{Christopher V. Poulton, Xiaoge Zeng, Mark T. Wade and Milo\v{s} A. Popovi\'{c}$^{*}$}

\address{Department of Electrical, Computer, and Energy Engineering, University of Colorado, Boulder, CO  80309, USA}

\email{$^*$milos.popovic@colorado.edu}

\begin{abstract}We demonstrate an add-drop filter based on a dual photonic crystal nanobeam cavity system that emulates the operation of a traveling-wave resonator and drops light on resonance to a single output port. Realized on an advanced SOI CMOS (IBM 45\,nm SOI) chip without any foundry process modifications, the device shows 16\,dB extinction in through port and 1\,dB loss in drop port with a 3\,dB bandwidth of 64\,GHz. To the best of our knowledge, this is the first implementation of a four-port add-drop filter based on photonic crystal nanobeam cavities.


\end{abstract}

\ocis{(230.5298) Photonic crystals; (130.7408) Wavelength filtering devices; (200.4650) Optical interconnects.} 

]

\noindent 
Wavelength division multiplexing (WDM) has become a promising scheme for high-capacity optical interconnect and communication\cite{WDM1}. Channel add-drop optical filter is a critical component for WDM, and has been  
previously demonstrated in microring resonators\cite{WDM1, OrcuttPlat} whose traveling-wave mode structure enables complete separation of input, drop and through ports without the need for optical circulators.  Photonic crystal (PhC) nanobeam cavities\cite{LoncarHighQPhC,LoncarPhCDesign} have tight optical mode confinement and high quality factors enabling applications in many photonic devices including lasers, sensors, nonlinear and opto-mechanical devices\cite{LoncarPhCThermal,PHCApplication,PhCNonlinear,PhCOptomechanics}. However, their usage in add-drop filters is non-trivial due to the fact that a standing-wave resonator couples light into traveling-wave modes of opposite propagation directions. In a previous demonstration of wavelength filtering in a PhC nanobeam cavity, 
a circulator \cite{LoncarFeedingWGPhC,LoncarPhCThermal} is required to route reflected wave into through port, making it impractical for usage in cascaded channel add-drop filters. 
Although cascaded filters for WDM applications have been proposed \cite{Noda2DPhCFilter1} and realized \cite{Noda2DPhCFilter2} in two-dimensional photonic crystal slabs, this devices are bulky and rely on wavelength-sensitive heterostructure interfaces. 
Over 15 years ago, Manolatou et al. proposed that a channel add-drop filter can be implemented in a pair of coupled standing-wave resonators \cite{manolatou}, however, to date, this concept has not been demonstrated.


In this Letter, we demonstrate an efficient channel add-drop filter based on a pair of photonic crystal nanobeam cavities, demonstrating for the first time a high-performance 4-port PhC nanobeam standing-wave resonator based filter directly integrable into an on-chip WDM scheme without any magneto-optic components.  The filter has 1\,dB insertion loss and 16\,dB through port extinction for a 3\,dB bandwidth of 64\,GHz.  We demonstrate the device in an unmodified 45\,nm SOI CMOS process, on a chip fabricated in a commercial microelectronics foundry\cite{OrcuttInte,OrcuttPlat}.  This work thus lays the groundwork for greater utilization of photonic crystal devices in complex photonic chips.  The cavities are not tuned, and the fidelity of the process is shown to be high enough to enable degenerate operation of the cavity pair.

\begin{figure}[b] 
  \centering
  \includegraphics[width=1\columnwidth]{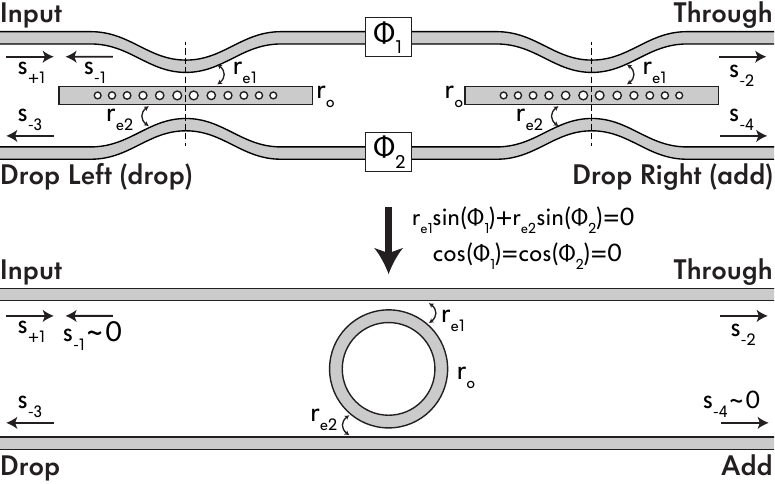}
\caption{Topology of device\cite{manolatou}. Two identical photonic crystal microcavities are evanescently coupled to two bus waveguides. To emulate a traveling-wave cavity, maximum add port suppression (in drop right) and minimum input reflection, $\cos(\phi_1) = \cos(\phi_2) = 0$ and $r_{e1}\sin(\phi_1)+r_{e2}\sin(\phi_2) = 0$, where $\phi_1$ and $\phi_2$ are the phases between the cavities' symmetry planes.\label{fig:topology}}
\end{figure}


Fig.~\ref{fig:topology} shows the device topology. Two photonic crystal nanobeams are indirectly coupled via two evanescently-coupled buses with no mutual coupling between the cavities. The system is excited on the left by a single input $s_{+1}$. $\phi_1$ and $\phi_2$ are the optical phase delays in the two connecting paths between the two cavities' symmetry planes. The advantage of this topology is that when $\phi_1$ and $\phi_2$ are engineered correctly, destructive interference occurs in one of the drop ports and the input port, and only a single drop port and the through port will have non-zero transmission for all wavelengths. On resonance, one of the drop ports will have near $100\%$ transmission, limited only by the loss of the system. In a single evanescently coupled photonic crystal cavity, all four ports will have non-zero transmission on-resonance, allowing for a maximum of $25\%$ from any single drop port owing to symmetry restrictions. 

In order to analyze the effect of the cavities not having an identical resonant frequency, the coupled mode theory model of the device derived in \cite{manolatou} was expanded to include a resonant frequency difference between the two cavities and is 

\vspace{-12pt}
\begin{multline}
\frac{da_L}{dt}=(j(\omega_o+\delta\omega)-r_{e1}-r_{e2}-r_o)a_L\\
+\sqrt{r_{e1}}e^{j\theta}s_{+1}-r_{e1} e^{-\phi_1}a_R-r_{e2}e^{-\phi_2}a_R
\end{multline}
\vspace{-22pt}
\begin{multline}
\frac{da_R}{dt}=(j(\omega_o-\delta\omega)-r_{e1}-r_{e2}-r_o)a_R\\
+\sqrt{r_{e1}}e^{j\theta}e^{-\phi_1}s_{+1}-r_{e1} e^{-\phi_1}a_L-r_{e2}e^{-\phi_2}a_L
\end{multline}

\noindent where $r_{e1}$ is the decay rate to the bus, $r_{e2}$ is the decay rate to the receiver, $r_o$ is the decay rate due to loss in each cavity, $\omega_o$ is the average of the two cavities resonant frequencies, $\delta\omega$ is half of the resonant frequency difference between the two cavities, and $\theta$ is the phase of the coupling coefficient from both the bus and the receiver. All decay rates are related to the intrinsic/unloaded Q and the two external/coupling Q's by $r_o=\omega_o/(2Q_o)$, $r_{e1}=\omega_o/(2Q_{e1})$, and $r_{e2}=\omega_o/(2Q_{e2})$, where $\delta\omega$ is assumed to be small compared to $\omega_o$. For the device to act as a traveling-wave cavity, the phases $\phi_1$ and $\phi_2$ are set so that the symmetric and anti-symmetric supermodes of the device are degenerate. For degenerate cavities, this occurs when $r_{e1}\sin(\phi_1)+r_{e2}\sin(\phi_2) = 0$ and  $\cos(\phi_1) = \cos(\phi_2) = 0$ \cite{manolatou} and is assumed to be similar for this system. Therefore, not only the relative difference of the two phases must be set but also the absolute value of each individual phase. The left drop port is the active drop port when this condition is met and the right drop port acts as an add port. The field outputs of the through port and the left and right drop ports of the device at the assumed degeneracy condition for our model at $\omega_o$ are:

\begin{align}
\frac{s_{-2}}{s_{+1}} &= -i\frac{\delta\omega^2+2r_{e2}(r_o+r_{e2}-r_{e1})+r_o^2}{\delta\omega^2+2(r_{e1}^2+r_{e2}^2)+2r_o(r_{e1}+r_{e2})+r_o^2}\label{equ:through}\\
\frac{s_{-3}}{s_{+1}} &= -\frac{2\sqrt{r_{e1}r_{e2}}(r_o+r_{e1}+r_{e2})}{\delta\omega^2+2(r_{e1}^2+r_{e2}^2)+2r_o(r_{e1}+r_{e2})+r_o^2}\label{equ:leftDrop}\\
\frac{s_{-4}}{s_{+1}} &= \frac{2\sqrt{r_{e1}r_{e2}}(\delta\omega+i(r_{e2}-r_{e1}))}{\delta\omega^2+2(r_{e1}^2+r_{e2}^2)+2r_o(r_{e1}+r_{e2})+r_o^2}\label{equ:rightDrop}
\end{align}

\begin{figure}[t!]
  \centering
  \includegraphics[width=.8\columnwidth]{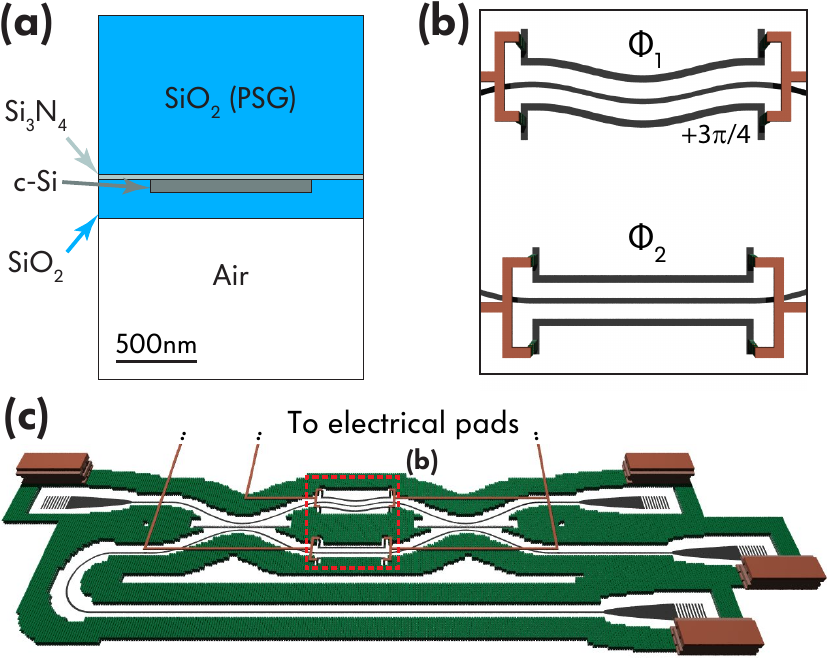}
\caption{(a) Cross-section of cavity in IBM 45\,nm 12SOI CMOS (see PDK \cite{ibm12soi}); (b) Relative phase offset of about $3\pi/4$ realized with a sinusoid waveguide; (c) 3D rendering of device mask-set layout. \label{fig:design}}
\end{figure}

\noindent Eq.~(\ref{equ:leftDrop}) shows that the drop loss in the left drop port is not affected greatly by a small resonant frequency difference between the cavities as both $r_{e1}^2$ and $r_{e2}^2$ are expected to be large compared to $\delta\omega^2$ in the denominator. However, Eq.~(\ref{equ:rightDrop}) indicates that the suppression in the right drop port is completely limited by the detuning between the two cavities along with the difference of coupling between the bus and receiver because $\Re(\frac{s_{-4}}{s_{+1}})\propto\delta\omega$ and $\Im(\frac{s_{-4}}{s_{+1}})\propto(r_{e2}-r_{e1})$. If low drop loss is the main requirement of the filters then as long as the fabrication process has decent fidelity, then individual cavity tuning may not be necessary (and is not used in the device presented here). Note these equations and analysis is true for this device topology with any standing-wave cavities and not only photonic crystals.

\begin{figure}[t!]
  \centering
  \includegraphics[width=.8\columnwidth]{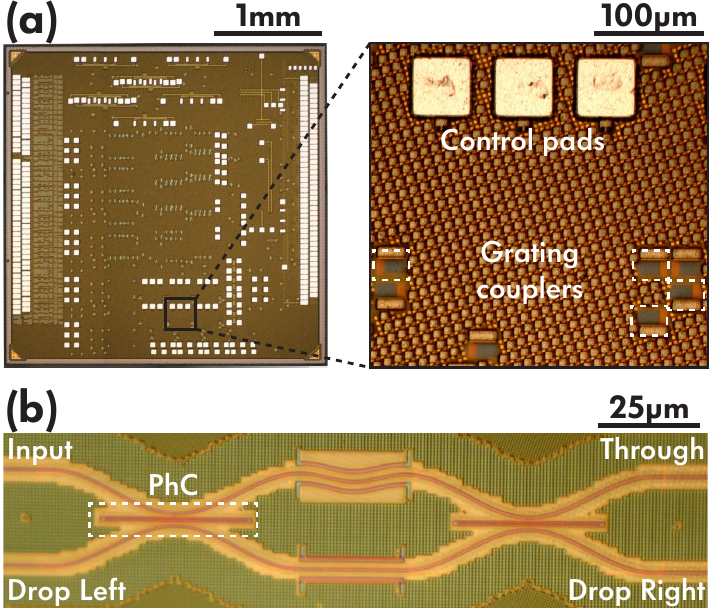}
\caption{(a) Optical micrographs of fabricated device from the front and (b) back of chip (substrate removed).\label{fig:device}}
\end{figure}

\begin{figure*}[t!] 
  \centering
  \includegraphics[width=2\columnwidth]{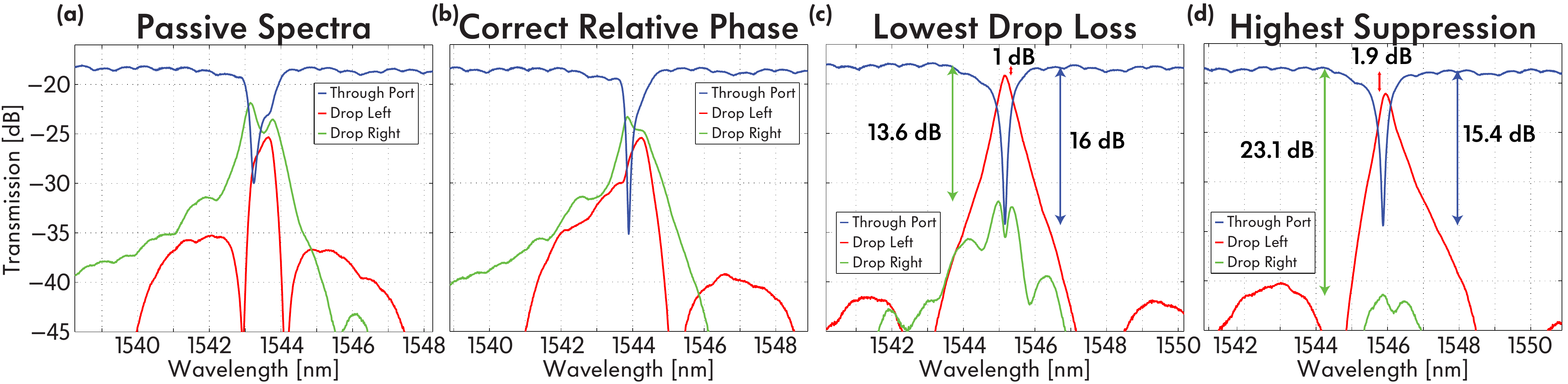}
\caption{(a) Spectra of device without any thermal phase tuning; (b) Spectra of device at estimated correct relative phase ($|\phi_1-\phi_2|=\pi$); (c) Spectra of device after tuning absolute phases for lowest drop loss and (d) highest add-port suppression.\label{fig:results}}
\end{figure*}

The device was fabricated in an unmodified commercial microelectronics 45\,nm SOI CMOS process, the IBM 12SOI process. The cross-section of the device is illustrated in Fig.~\ref{fig:design}(a). The main challenge of design is the sub-100\,nm device layer. Using similar designs of photonic crystal microcavities in \cite{PoultonCavs}, two cavities were cascaded. A sinusoidal waveguide on the top path was used to bias the relative phase between the top and bottom paths [Fig.~\ref{fig:design}(b)]. By biasing one of the arms, the phase tuning of the device can be simplified but is not absolutely necessary. If included, the phase of both arms can be tuned near equally to bring the phases to the correct absolute phase without having to adjust the relative phase difference. The additional phase gained in the top arm is near $3\pi/4$ for wavelengths around 1550\,nm which was the designed photonic crystal resonant frequency. A bias of $\pi$ was not chosen because of fabrication tolerances. Fig.~\ref{fig:design}(c) shows a 3D rendering of the device. Two highly-doped resistive heaters implemented in the crystalline silicon layer (the active layer for transistors) are placed next to each path to act as phase tuners. Both phase tuners are needed to bring both paths to the correct absolute phase and to fine tune the relative phase between the two paths. Fig.~\ref{fig:device}(a) is a top-view optical micrograph of a 3$\times$3\,mm die from a 300\,mm, 45\,nm node SOI CMOS wafer, and a zoom in view showing grating coupler access ports to the device (not visible, under metal density fill) and heater driving pads for driving probes. Fig.~\ref{fig:device}(b) is a bottom-view optical micrograph of a fabricated device (grating couplers not shown) after the silicon substrate is removed. 

The spectra of the fundamental resonance of the device without any thermal phase tuning is shown in Fig.~\ref{fig:results}(a). Because of the built in phase bias of the $\phi_1$ arm, the three ports of the device are not symmetrical. Fig.~\ref{fig:results}(b) shows the spectra of the device after tuning the $\phi_1$ arm to bring to the best performance possible in terms of through port extinction and in theory setting $|\phi_1-\phi_2|=\pi$. A more Lorentzian shaped through port is seen but the drop port is still asymmetrical and the drop loss is large. $\phi_1$ and $\phi_2$ were then tuned equally in additional power until the lowest drop loss was achieved. This spectra is shown in Fig.~\ref{fig:results}(c). For this tuning of $\phi_1$ and $\phi_2$, the device has a through port extinction of 16\,dB and an insertion loss of 1\,dB through the left drop port on-resonance. Transmission to the right drop port is still below $-13.6$\,dB, showing successful isolation of that port over the operational wavelength range. A bandwidth of 64\,GHz (a loaded Q of 3,020) is measured. Detuning the phases slightly away from this point allows for a much higher suppression with an increase of insertion loss and this is shown in Fig.~\ref{fig:results}(d). This is near the degeneracy condition because the device has the highest right drop port suppression seen which indicates a traveling-wave resonator like response and thus degenerate supermodes. The increase of drop loss is believed to be because of a slight difference of coupling between the left and right direction on the bus and/or receiver as it is not included in our model. The suppression of the right drop port is 21.3\,dB with a drop loss of 1.9\,dB and a through port extinction of 15.4\,dB. 


\begin{figure}[b!]
  \centering
  \includegraphics[width=.8\columnwidth]{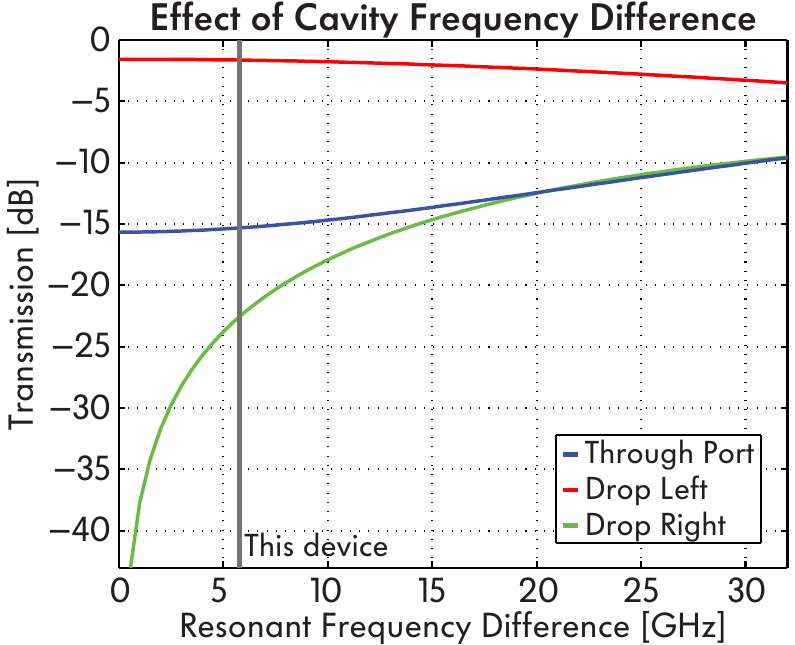}
\caption{Port power transmission at center frequency as a function of cavity resonant frequency difference with the extracted values of $r_o$=33.2\,Grad/s and $r_{e1}$=$r_{e2}$=84.2\,Grad/s ($Q_o$=18,350 and $Q_{e1}$=$Q_{e2}$=7,240).\label{fig:detuning}}
\end{figure}

The detuning between the two cavities and the internal and external Q's of the device can be estimated by assuming that the bus and receiver coupling rates are equal, and fitting the measured values for the insertion loss, through port extinction and the add port suppression to Eq.~\cref{equ:leftDrop,equ:rightDrop,equ:through} and utilizing the measured loaded Q. The total detuning between the two cavities is fitted to be 5.4\,GHz, the intrinsic Q of each cavity to be 18,350 and the external Q's to the bus and receiver to be 7,240 (corresponding to a $\delta\omega$ of 16.9\,Grad/s, a $r_o$ of 33.2\,Grad/s, and a $r_{e1}$ and a $r_{e2}$ of 84.2\,Grad/s). In order to see the benefit if thermal resonant frequency tuning were added to the photonic crystals, Fig.~\ref{fig:detuning} shows the theoretical performance of a device as a function of the difference of the resonant frequencies of the two cavities with $r_o$ and $r_e$ fixed to the estimated values. The drop loss and through port extinction is not greatly limited by the resonance frequency difference of the two cavities. Instead they are limited by the ratio of the intrinsic and external Q which can be improved by having a better photonic crystal design. For these cavities, up to a resonant frequency difference of 27\,GHz the drop loss is less than 3\,dB. However, the right drop port suppression decreases quickly with the resonant frequency difference. In theory, thermal tuning could have greatly increased the right drop suppression but any difference of the bus and receiver coupling could reduce this benefit as seen before in Eq.~(\ref{equ:rightDrop}). At resonance frequency differences greater than 32\,GHz, noticeable resonance splitting occurs in spectra of the left drop port around the center frequency.

The demonstration of cascaded photonic crystal standing-wave microcavities with a traveling-wave cavity (ring resonator) like response enables photonic crystals to be used in an add-drop filter configuration with low insertion loss and large extinction in monolithic CMOS photonic circuits. The device shown can be cascaded to create a WDM system that is comparable to systems previously realized with microring resonators because of its 4-port add-drop qualities. Though the cavities shown here are not tunable, tunable PhC nanobeams have been shown within this fabrication process\cite{PoultonTunableCavs} that could be used in order to tune the WDM channels as needed and to reduce the resonant frequency difference between the two microcavities, further increasing the drop port transmission, through port extinction, and add port suppression. A remaining challenge preventing wide adoption of such devices is the complex implementation.  A simplified implementation of the same concept, with fewer phase adjustments necessary, could enable wider adoption of PhC devices in photonic chips for WDM applications. 

This work was supported by DARPA POEM program award HR0011-11-C-0100.

\end{document}